\documentclass[10pt,twocolumn,superscriptaddress,floats,showpacs,nobalancelastpage,longbibliography,prb]{revtex4-2}

\usepackage[pdftex]{color}
\usepackage[dvipsnames]{xcolor}

\usepackage{array,longtable}
\newcolumntype{L}{>{\tiny $}p{0.33\columnwidth}<{$}}
\newcolumntype{M}{>{\scriptsize $}p{0.33\columnwidth}<{$}}
\newcolumntype{N}{>{\scriptsize $}p{0.43\columnwidth}<{$}}
\usepackage{amsmath}
\usepackage{amssymb}
\usepackage{amsfonts}
\usepackage{graphicx}
\usepackage{hyperref}
\usepackage{tabularx}
\usepackage{wasysym}
\usepackage{bbold}
\usepackage{orcidlink}

\usepackage{float}
\usepackage{graphics}
\usepackage[caption=false]{subfig}
\usepackage{tikz}
\usepackage{times}
\usepackage{dsfont}
\usepackage{braket}
\usepackage{subfig}

\newcommand{\CNOT}[1]{\mathcal{C\!\!\!/\,}^{#1}}
\newcommand{\Sgate}[1]{\mathcal{S}^{#1}}
\newcommand{\Hgate}[1]{\mathcal{H}^{#1}}
\newcommand{\mat}[1]{\left(\begin{matrix}#1\end{matrix}\right)}

\newcommand{\mycomment}[1]{}

\usepackage[normalem]{ulem}

\allowdisplaybreaks[1]

\newif\ifhyper
\hypertrue
\ifhyper
\hypersetup{
   citecolor = {OliveGreen},
   colorlinks = {true}, 
   urlcolor = {RoyalBlue}, 
   linkcolor = {RoyalBlue}
}
\fi

\begin{document}

\title{Clifford Circuits Augmented Grassmann Matrix Product States}

\author{Atis Yosprakob\,\orcidlink{0000-0001-8597-1460}}
\email{yosprakob@yukawa.kyoto-u.ac.jp}
\affiliation{Yukawa Institute for Theoretical Physics, Kyoto University
Kitashirakawa Oiwakecho, Sakyo-ku, Kyoto 606-8502 Japan}

\author{Wei-Lin Tu\,\orcidlink{0000-0002-3340-4963}}
\email{weilintu@keio.jp}
\affiliation{Graduate School of Science and Technology, Keio University, 3-14-1 Hiyoshi, Kohoku, Yokohama, Kanagawa 223-8522, Japan}
\affiliation{Keio University Sustainable Quantum Artificial Intelligence Center (KSQAIC), Keio University, Tokyo 108-8345, Japan}

\author{Tsuyoshi Okubo\,\orcidlink{0000-0003-4334-7293}}
\email{tsuyoshi.okubo.phys@niigata-u.ac.jp}
\affiliation{Department of Physics, Niigata University, Niigata, 950-2181, Japan}

\author{Kouichi Okunishi\,\orcidlink{0000-0002-4620-1241}}
\email{okunishi@omu.ac.jp}
\affiliation{Department of Physics, Graduate School of Science, Osaka Metropolitan University, Osaka 558-8585, Japan}
\affiliation{Nambu Yoichiro Institute of Theoretical and Experimental Physics (NITEP), Osaka Metropolitan University, 3-3-138 Sugimoto, Osaka 558-8585, Japan }

\author{Donghoon Kim\,\orcidlink{0009-0002-8358-5253}}
\email{donghoon.kim@riken.jp}
\affiliation{Analytical Quantum Complexity RIKEN Hakubi Research Team, RIKEN Center for Quantum Computing (RQC), Wako, Saitama 351-0198, Japan}

\date{\today}

\begin{abstract}

Recent progress in combining Clifford circuits with tensor-network (TN) methods has shown that local Clifford disentanglers can reduce bipartite entanglement across TN bonds prior to tensor compression, thereby improving the efficiency of TN simulations.
In this work, we embed local Clifford disentanglers in the Grassmann-tensor language to define a Clifford-augmented Grassmann matrix product state (CAGMPS) ansatz, and develop a density-matrix renormalization group (DMRG) framework based on this ansatz while preserving locality and fermion-parity structure.
We benchmark the resulting CAGMPS--DMRG method on representative fermionic lattice systems, including the tight-binding, $t$--$V$, and $t$--$V$--$V'$ models.
In all cases, Clifford augmentation systematically suppresses bipartite entanglement and improves the accuracy of the ground-state energy at a fixed bond dimension.
We further show that the Grassmann-evenness condition, together with equivalence under entangling action, restricts the relevant two-site Clifford candidates to 12 inequivalent representatives, enabling a more economical disentangling search than approaches based on the standard two-qubit Clifford gate set.
Our results suggest that the CAGMPS--DMRG method provides a scalable and efficient variational tool for strongly correlated fermionic systems.

\end{abstract}

\maketitle

\section{Introduction}
\label{sec:Introduction}

Strongly correlated fermionic systems provide a fundamental setting for understanding a wide range of emergent quantum phenomena, including high-temperature superconductivity, Mott transitions, and non-Fermi-liquid behavior~\cite{Anderson1987,Dagotto1994,Georges1996,Imada1998,Lee2006,Sachdev2011,Keimer2015,Qin2022,Arovas2022}.
Despite their broad relevance, theoretical and numerical treatments of such systems remain notoriously difficult, owing to the combined effects of strong correlations and fermionic statistics, which often lead to severe computational challenges, such as the sign problem in Monte Carlo approaches~\cite{Loh1990, Chandrasekharan1999, Troyer2005}.
Consequently, the development of accurate, scalable, and systematically improvable numerical approaches for interacting fermions remains a central goal in many-body physics.

Among these approaches, tensor-network (TN) methods based on matrix product states (MPS)~\cite{Ostlund1995,Verstraete2008,Cirac2009,Schollwock2011,Orus2014,Cirac2021}, most notably the density-matrix renormalization group (DMRG)~\cite{White1992,White1993,Schollwock2005}, have been highly successful for one-dimensional (1D) quantum many-body systems.
When applied to fermionic models, however, these methods require special care in handling fermionic anticommutation relations, which are often encoded by mapping fermions to spin degrees of freedom via the Jordan--Wigner (JW) transformation~\cite{Jordan1928,Lieb1961,Bravyi2002,Verstraete2005mapping}.
In this representation, fermionic anticommutation is captured by JW strings, whose nonlocal support can obscure operator locality and becomes especially cumbersome in higher-dimensional geometries.
This motivates the use of fermionic TN formulations, which treat fermionic statistics directly while preserving locality~\cite{Barthel2009contraction,Corboz2009fermionic,Pineda2010unitary,Kraus2010fermionic,Corboz2010simulation,Bruognolo2021beginner}.

Another bottleneck of TN methods arises from entanglement.
Standard TN ansatzes are efficient when the target state has limited bond entanglement, as in many area-law states~\cite{Cirac2021MPSPEPS,Hastings2007AreaLaw,Schuch2008EntropyScaling,Arad2012FFAreaLaw,Arad2013AreaLaw,Brandao2013AreaLaw,Kuwahara2020LongRangeAreaLaw,Anshu2022AreaLaw2D,Kim2024FullyConnectedAreaLaw,Kim2024BosonAreaLaw}.
However, their efficiency deteriorates for highly entangled states, including critical systems and fermionic systems with extended gapless structures such as Fermi surfaces~\cite{Wolf2006AreaLawFermions,Gioev2006Widom,Swingle2010FermiSurface}.
A promising strategy is to apply local unitary transformations that reorganize the target wavefunction into a less entangled form before representing it with a TN~\cite{Vidal2007ER,Vidal2008MERA,Corboz2009FermionicMERA,Evenbly2015TNR}.
The effectiveness of this strategy, however, depends crucially on finding disentanglers that are expressive enough to capture dominant entanglement patterns while remaining compatible with efficient TN manipulations.

Recent works have developed Clifford-assisted TN methods that use Clifford circuits as such disentanglers~\cite{Qian2024, Qian2025,Qian2024temp,Fan2025disentangling,Huang2025nonstabilizerness,Huang2025fermion,Huang2025augmenting}.
Motivated by the Gottesman--Knill theorem~\cite{gottesman1998heisenberg}, this choice exploits the fact that Clifford circuits can generate highly entangled stabilizer states while remaining efficiently classically simulable.
This observation motivates separating Clifford-representable contributions to entanglement from the residual correlations encoded in TN tensors.
In practice, this separation is implemented by applying Clifford circuits before tensor compression to reduce the entanglement carried by TN bonds.
This hybrid strategy has been explored in several spin and qubit-based settings~\cite{Tagliacozzo2011entanglement,mello2024hybrid,mello2025clifford,Frau2025stabilizer,Mello2025Loschmidt,Llima2024,Nakhl2025,Collura2024tensor}.
It has also been applied to fermionic systems after mapping fermions to qubits~\cite{Huang2025fermion}.
However, as discussed above, fermion-to-qubit mappings introduce mapping-dependent nonlocal strings, which can obscure how qubit Clifford disentanglers relate to the locality structure of the original fermionic problem.
These considerations underscore the need for a formulation that incorporates Clifford disentanglers directly into fermionic TNs while preserving this locality structure.

In this work, we develop a Clifford-assisted Grassmann TN framework for DMRG simulations of fermionic systems.
We define local Clifford circuits directly in the Grassmann formalism and incorporate them into Grassmann tensor networks (GTNs) as variational disentanglers.
GTNs encode fermionic degrees of freedom using Grassmann variables~\cite{Gu2010grassmann,Gu2013efficient,Yosprakob2023grassmanntn}, whose anticommutation relations naturally reproduce fermionic statistics.
As a result, both the TN representation and the Clifford transformations act within the fermionic algebra, avoiding the nonlocal JW strings that arise in qubit-mapped formulations.
We formulate this construction as a Clifford-augmented Grassmann matrix product state (CAGMPS) ansatz within a DMRG framework, and benchmark the resulting method on the tight-binding, $t$--$V$, and $t$--$V$--$V'$ spinless fermion models.
Across these benchmarks, CAGMPS improves the accuracy of the ground-state energy at a fixed bond dimension and reduces the bipartite entanglement carried by MPS bonds.

This direct fermionic formulation has two main advantages.
First, since the Clifford transformations are defined within the Grassmann tensor formalism, the disentangling step acts locally on fermionic modes while preserving their anticommutation relations.
Second, and more importantly for efficiency, fermion-parity symmetry strongly constrains the Clifford disentangling step.
Physical fermionic operations must preserve fermion parity, which in the Grassmann formulation corresponds to restricting the local Clifford transformations to Grassmann-even operators.
After further identifying gates that are equivalent in their entangling action, the relevant two-site Clifford search space is reduced to 12 inequivalent representatives.
This compact parity-preserving gate set leads to a more efficient disentangling search than in qubit-mapped fermionic formulations based on the standard two-qubit Clifford gate set.

The remainder of this paper is organized as follows.
In Sec.~\ref{sec:GrassmannTN}, we review the Grassmann tensor formalism and introduce the Grassmann MPS (GMPS) representation used throughout this work.
In Sec.~\ref{sec:GrassmannClifford}, we define Grassmann Pauli operators and construct parity-preserving Grassmann Clifford circuits, showing how the Grassmann-evenness condition reduces the relevant two-site Clifford gate set.
In Sec.~\ref {sec:CAGMPS}, we present the DMRG algorithm based on the CAGMPS ansatz, including the local disentangling search and the corresponding transformation of the Hamiltonian.
In Sec.~\ref{sec:Benchmarks}, we benchmark the method on the tight-binding, $t$--$V$, and $t$--$V$--$V'$ spinless fermion models and compare its accuracy and entanglement structure with conventional GMPS.

\section{Grassmann Tensors and MPS}
\label{sec:GrassmannTN}

Let $\{\theta_a\}$ and $\{\theta^\dagger_a\}$ denote Grassmann generators and their duals, obeying the canonical anticommutation relations $\theta_a \theta_b = - \theta_b \theta_a$, $\theta^\dagger_a \theta^\dagger_b = - \theta^\dagger_b \theta^\dagger_a$, and $\theta^\dagger_a \theta_b = - \theta_b \theta^\dagger_a$. For a fermionic leg $\psi$, we write monomials $\psi^I \equiv \theta_1^{i_1}\theta_2^{i_2}\cdots$ with a composite index $I=(i_1,i_2,\dots)$ and Grassmann parity $p(I)=\sum_k i_k \ (\mathrm{mod}\ 2)$. A Grassmann tensor with $m$ non-conjugated legs $\{\psi_a\}_{a=1}^m$ and $n$ conjugated legs $\{\phi^\dagger_b\}_{b=1}^n$ (order $(m,n)$) is defined as a linear combination of Grassmann monomials with complex coefficients:
\begin{align}
&T_{\psi_1\cdots\psi_m\,\phi^\dagger_1\cdots\phi^\dagger_n}\nonumber
\\&= \sum_{I_1,\dots,I_m,\,J_1,\dots,J_n}
T_{I_1\cdots I_m J_1\cdots J_n}\;
\psi_1^{I_1}\cdots\psi_m^{I_m}\,
\phi_1^{\dagger J_1}\cdots\phi_n^{\dagger J_n}.
\label{eq:grassmann_tensor_general}
\end{align}
By changing the ordering of the monomials, i.e., the tensor signature, one gets different tensor coefficients. We therefore should fix the signature of each Grassmann tensor during the calculation.

Contractions are defined only between dual pairs of generators and are implemented by Berezin integration with a Gaussian kernel. For a dual pair $\psi^I=\theta_1^{i_1}\cdots\theta_n^{i_n}$ and $\psi^{\dagger J}=\theta_n^{\dagger j_n}\cdots\theta_1^{\dagger j_1}$, the orthogonality relation is
\begin{equation}
\label{eq:grassmann_contraction_rule}
\int_{\psi^\dagger,\psi}\; \psi^{I}\,\psi^{\dagger J}
= \delta_{IJ}.
\end{equation}
with $\int_{\psi^\dagger,\psi}F:=\int \prod_{a=1}^n d\theta^\dagger_a\, d\theta_a\; e^{-\theta^\dagger_a \theta_a}F$.
For two tensors $A_{\psi\,\cdots}$ and $B_{\cdots\,\psi^\dagger}$, contracting the shared leg yields
\begin{align}
    \big(A \star_{\psi^\dagger,\psi} B\big)_{\cdots} &= 
    \int_{\psi^\dagger,\psi}\; A_{\psi\,\cdots}\, B_{\cdots\,\psi^\dagger} \\
    &= \sum_{I} s_{I}\, A_{I\,\cdots}\, B_{\cdots\, I},
\end{align}
where the ellipses denote the remaining (uncontracted) legs and their composite indices, and $s_I$ denotes the sign factor arising from rearranging the signature into the contractable form \eqref{eq:grassmann_contraction_rule}. This algebraic rule is the only nontrivial ingredient needed to compose large fermionic networks while preserving exact antisymmetry.

Consider an $n$-site fermionic wave function $|\Phi\rangle = \sum_{i_1,\dots,i_n} A_{i_1\cdots i_n}\,|i_1\rangle\otimes\cdots\otimes|i_n\rangle$, with local physical dimension $D$. 
A GMPS factorizes the coefficient tensor into a chain of local Grassmann tensors, each with one physical index $\psi_a$ and two virtual indices $(\phi_{a-1},\phi^\dagger_a)$ with bond dimension $\chi$:
\begin{align}
A_{\psi_1\cdots \psi_n}
&=\sum_{i_1,...,i_n}A_{i_1\cdots i_n}\psi_1^{i_1}\cdots \psi_n^{i_n}\\
&=
\prod_{a=1}^{n-1} \int_{\phi^\dagger_a,\phi_a}
\Big(M^{(1)}_{\,\psi_1\,\phi_1}
M^{(2)}_{\phi^\dagger_1\, \psi_2\, \phi_2}
\!\!\!\!\!\!\!\cdots
M^{(n)}_{\phi^\dagger_{n-1}\, \psi_n}\Big),
\label{eq:gmps_definition}
\end{align}
where each site tensor admits the component form of \eqref{eq:grassmann_tensor_general}, e.g.
$M^{(a)}_{\phi^\dagger_{a-1}\, \psi_a\, \phi_a}
=\sum_{I,J} M^{(a)}_{I\, i\, J}\, \phi_{a-1}^{\dagger I}\psi^i_a \phi_a^{J}$.
All nearest-neighbor contractions in \eqref{eq:gmps_definition} reduce, via \eqref{eq:grassmann_contraction_rule}, to finite sums over composite indices. Thus the GMPS inherits the structural efficiency of conventional MPS (parameter count $\mathcal{O}(n D \chi^2)$) while enforcing fermionic antisymmetry algebraically, without JW strings or swap gates. In practice, one chooses site tensors to be Grassmann-even (each monomial carries net even fermion number) so that the ansatz respects the fermion-parity superselection rule and interfaces cleanly with parity-preserving variational updates (e.g.\ DMRG).
All calculations involving Grassmann tensors in this work were carried out with the open-source Python package GrassmannTN \cite{Yosprakob2023grassmanntn}.

\begin{figure*}[t]
    \centering
    \includegraphics[width=\linewidth]{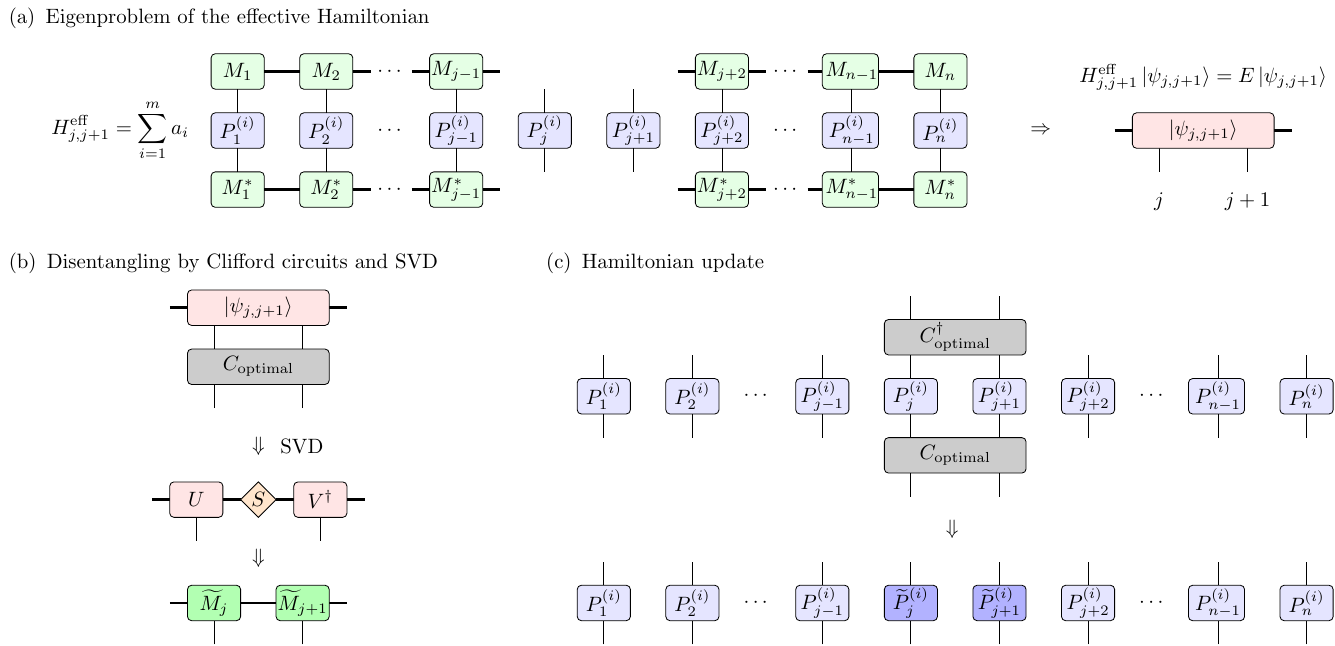}
    \caption{Overview of the CAGMPS-based DMRG algorithm.
    (a) As in conventional two-site DMRG, to update the GMPS on two adjacent sites $j$ and $j+1$, one constructs the effective Hamiltonian from the given GMPS, where the Hamiltonian is now represented as Grassmann Pauli strings, and then solves its eigenproblem to obtain the eigenvector $\ket{\psi_{j,j+1}}$.
    (b) The state $\ket{\psi_{j,j+1}}$ is disentangled across sites $j$ and $j+1$ using a Grassmann Clifford circuit. Among all possible Grassmann Clifford circuits, the optimal one that minimizes the entanglement is then identified and applied, after which the updated GMPS is obtained via singular value decomposition.
    (c) The same Grassmann Clifford circuit is also applied to the Hamiltonian. Due to the properties of Grassmann Clifford circuits, the Grassmann Pauli operators on sites $j$ and $j+1$ are finally mapped to other Grassmann Pauli operators.}
    \label{fig:algorithm}
\end{figure*}

\section{Grassmann Pauli operators and Clifford circuits} \label{sec:GrassmannClifford}

We define Grassmann counterparts of Pauli matrices as follows
\begin{equation}
    \varsigma^\mu_{\phi^\dagger\psi}=\sum_{ij}\sigma^\mu_{ij}\phi^{\dagger i}\psi^j
\end{equation}
where $\sigma^{\mu}$ with $\mu=x,y,z$ are the standard Pauli matrices and $\sigma^\text{id}=\mathbb{1}$. Analogously, fermionic creation and annihilation operators and the number operator can be written in terms of these Pauli matrices as
\begin{align}
    c&=\frac{1}{2}(\varsigma^x+i\varsigma^y),\\
    c^\dagger&=\frac{1}{2}(\varsigma^x-i\varsigma^y),\\
    n&=\frac{1}{2}(\varsigma^\text{id}-\varsigma^z).
\end{align}
A Grassmann Clifford circuit is a unitary transformation $U$ that preserves the set of Pauli operators
\begin{equation}
     C P C^\dagger \in \mathcal{P}, \qquad \forall P \in \mathcal{P}.
\end{equation}
This is the direct Grassmann analog of the spin-$\frac12$ Clifford-group closure property.

To construct a set of Clifford gates, a sequence of operators from the set $\{\Hgate{0},\Sgate{0},\Hgate{1},\Sgate{1},\CNOT{01},\CNOT{10}\}$, the Grassmann counterparts of the Hadamard, $S$, and CNOT gates, is applied to the two-qubit identity operator
\begin{equation}
    \mathcal{I}_{\phi^\dagger_1\phi^\dagger_2\psi_2\psi_1}=\varsigma^\text{id}_{\phi^\dagger_1\psi_1}\varsigma^\text{id}_{\phi^\dagger_2\psi_2}
\end{equation}
until we obtain no further results. This amounts to all 11,520 elements of the two-qubit Clifford group in the Grassmann representation. 
We then further reduce this number to 720 by quotienting out the Pauli group, i.e., imposing the sign-positivity condition
\begin{align}
    CPC^\dagger &= +\varsigma^{\mu} \otimes \varsigma^{\nu},
\end{align}
for all $P\in\{\varsigma^x\otimes\varsigma^\text{id}, \varsigma^z\otimes\varsigma^\text{id}, \varsigma^\text{id}\otimes\varsigma^x, \varsigma^\text{id}\otimes\varsigma^z\}$ and for some $\mu,\nu \in \{\text{id},x,y,z\}$.

Additionally, we impose the Grassmann-evenness condition on the Grassmann Clifford gates. An operator is Grassmann-even if it contains an even number of Grassmann generators $(\theta_a, {\theta}^\dagger_a)$ in every term of its expansion. Grassmann-even Clifford circuits commute with the total fermionic parity operator
\begin{equation}
    \mathcal{P}_f = \bigotimes_{a\in \text{modes}} \varsigma^z_a,
\end{equation}
ensuring that they act block-diagonally in the even/odd fermionic parity sectors, consistent with physical fermionic evolutions. This condition further reduces the number of Clifford gates to 32. Finally, we quotient out redundancies under the left action of single-site gates, yielding 12 inequivalent gates, listed in Appendix~\ref{App:List}.

\section{Clifford-augmented GMPS: Variational Algorithm}\label{sec:CAGMPS}

Our algorithm follows the structure of the standard two-site DMRG, augmented with a disentangling step implemented by Clifford circuits as proposed in Ref.~\cite{Qian2024}. The key differences are the replacement of the conventional MPS with a GMPS and the tensor contractions that inherently account for fermionic statistics.
A general Hamiltonian for fermionic systems with $n$ sites can be expressed as a sum over Grassmann Pauli strings,
\begin{align}
    H = \sum_{i=1}^{m} a_{i} P^{(i)}, 
    \quad 
    P^{(i)} = P_{1}^{(i)} \otimes P_{2}^{(i)} \otimes \cdots \otimes P_{n}^{(i)},
\end{align}
where $a_{i} \in \mathbb{C}$ are constants, and $P^{(i)}$ denotes a Grassmann Pauli string with the $j$th site operator $P_{j}^{(i)} \in \{\mathds{1}, \varsigma^{x}, \varsigma^{y}, \varsigma^{z}\}$.

We now focus on two adjacent sites, $j$ and $j+1$, out of the total $n$ sites. The goal is to update the GMPS associated with these sites using the current GMPS as input. As illustrated in Fig.~\ref{fig:algorithm}(a), we first contract the GMPS with the Grassmann Pauli strings $P^{(i)}$ and sum the resulting contributions to construct the effective Hamiltonian $H_{j,j+1}^{\mathrm{eff}}$. Since we aim to approximate the ground-state GMPS, we update the local GMPS tensors by replacing their two-site wavefunction with the ground state $\ket{\psi_{j,j+1}}$ of $H_{j,j+1}^{\mathrm{eff}}$. This yields an updated GMPS that provides a better approximation to the true ground state.

Up to this point, the procedure corresponds to a Grassmann extension of the two-site DMRG. To further improve the approximation, we adopt the disentangling strategy of Ref.~\cite{Qian2024}, which employs Clifford circuits to remove classically simulable entanglement. Fixing the two-site wavefunction $\ket{\psi_{j,j+1}}$, we apply each candidate two-site Grassmann Clifford gate and evaluate the entanglement between sites $j$ and $j+1$. Among these candidates, we select the optimal gate, $C_{\mathrm{optimal}}$, that minimizes the entanglement, thereby yielding the disentangled state $C_{\mathrm{optimal}}\ket{\psi_{j,j+1}}$, as illustrated in Fig.~\ref{fig:algorithm}(b). The disentangled state is subsequently subjected to singular value decomposition (SVD), from which we obtain the updated GMPS tensors $\widetilde{M}_{j}$ and $\widetilde{M}_{j+1}$.

Since the application of $C_{\mathrm{optimal}}$ to the state corresponds to a local unitary transformation, the Hamiltonian must also be transformed accordingly as $H \rightarrow C_{\mathrm{optimal}} H C_{\mathrm{optimal}}^{\dagger}$,
as depicted in Fig.~\ref{fig:algorithm}(c). The Grassmann Clifford gate maps each Grassmann Pauli string to another Grassmann Pauli string. Consequently, the local components $P_{j}^{(i)}$ and $P_{j+1}^{(i)}$ of the Grassmann Pauli strings $P^{(i)}$ are transformed into $\widetilde{P}_{j}^{(i)}$ and $\widetilde{P}_{j+1}^{(i)}$ under $C_{\mathrm{optimal}}$, namely,
\begin{align}
    \widetilde{P}_{j}^{(i)} \otimes \widetilde{P}_{j+1}^{(i)} 
    = C_{\mathrm{optimal}} \, \big(P_{j}^{(i)} \otimes P_{j+1}^{(i)}\big) \, C_{\mathrm{optimal}}^{\dagger}.
\end{align}
This transformation can be implemented very efficiently within the algorithmic framework.

\section{Benchmarks}\label{sec:Benchmarks}

We benchmark the proposed CAGMPS-based DMRG algorithm against conventional GMPS-based DMRG on representative fermionic lattice models.
We focus on the ground-state energy error, the bipartite entanglement entropy, and the finite-size scaling of the entanglement entropy.
These benchmarks allow us to assess both the numerical accuracy and the compression efficiency gained by the Clifford augmentation.

\begin{figure}[t]
    \centering
    \includegraphics[width=\linewidth]{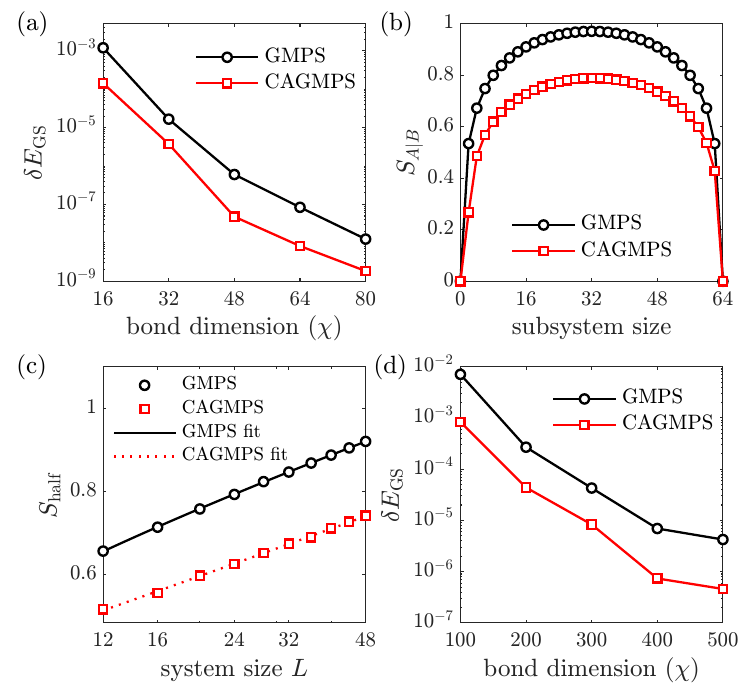}
    \caption{Benchmarks for the tight-binding model.
    (a) Ground-state energy error of GMPS and CAGMPS for the 1D tight-binding model in Eq.~\eqref{eq:TB_model}, plotted against the bond dimension $\chi$ for $L=64$.
    (b) Bipartite entanglement entropy of the same model, resolved along the chain by the cut position, for $L = 64$ and $\chi = 80$.
    (c) System-size scaling of the central-cut entanglement entropy $S_{\mathrm{half}}$, where the chain is bipartitioned into two halves, for the 1D tight-binding model at $\chi=64$. 
    The solid and dotted curves denote fits to $f(L)= (1/6) \log L + a + b/L$ for GMPS and CAGMPS, respectively.
    The fitted parameters are $a = 0.286$, $b = -0.528$ for GMPS and $a = 0.0938$, $b = 0.0765$ for CAGMPS.
    (d) Ground-state energy errors of GMPS and CAGMPS for the 2D tight-binding model on a $6 \times 6$ square lattice with increasing bond dimension $\chi$.}
    \label{fig:TB}
\end{figure}

We begin with the tight-binding Hamiltonian, 
\begin{align}
    H_{\text{TB}} &= - t \sum_{\langle i,j \rangle} \left( c_i^\dagger c_j + c_j^\dagger c_i \right), \label{eq:TB_model} 
\end{align}
where $\langle i,j \rangle$ denotes nearest-neighbor pairs on the underlying lattice, and $c_{i}^{\dagger}$ and $c_{i}$ are fermionic creation and annihilation operators.
We set $t = 1$ in the following calculations. 

We first consider the 1D version of Eq.~\eqref{eq:TB_model}.
Figure~\ref{fig:TB}(a) compares the ground-state energy errors of GMPS and CAGMPS for a chain of length $L = 64$.
CAGMPS achieves smaller errors at fixed bond dimension over the range of $\chi$ considered, demonstrating that Clifford augmentation improves the variational efficiency of the GMPS representation.
Figure~\ref{fig:TB}(b) shows the corresponding bipartite entanglement entropy for $L = 64$ and $\chi = 80$.
CAGMPS exhibits a lower entanglement profile than GMPS across the chain, indicating that the optimized Grassmann Clifford circuits reduce the entanglement carried by the MPS bonds.

For the same 1D tight-binding model, we further examine the system-size scaling of the bipartite entanglement entropy across the central cut.
We denote this half-chain entropy by $S_{\mathrm{half}}$.
For a critical free-fermion chain with open boundary conditions, conformal field theory predicts 
\begin{equation}
    S_{\mathrm{half}} = \frac{c}{6}\log L+S_0,
    \label{eq:EE_volume_scaling}
\end{equation}
with central charge $c = 1$.
As shown in Fig.~\ref{fig:TB}(c), both GMPS and CAGMPS follow this expected logarithmic scaling.
We fit the data using $f(L) = (1 / 6) \log L + a + b / L$, where the $1 / L$ term captures leading finite-size corrections.
The CAGMPS data retain the same universal scaling while exhibiting a smaller non-universal offset, consistent with the reduced entanglement observed in Fig.~\ref{fig:TB}(b).

We then apply the same tight-binding Hamiltonian (Eq.~\eqref{eq:TB_model}) to a two-dimensional (2D) $6 \times 6$ square lattice.
As shown in Fig.~\ref{fig:TB}(d), CAGMPS again achieves smaller energy errors than GMPS at the same bond dimension.
Although the 2D system is represented using a 1D MPS ordering, the result suggests that Clifford augmentation remains beneficial in higher-dimensional geometries.

Next, we consider the interacting spinless fermion $t$--$V$ model,
\begin{align}
    H_{t\text{--}V} &= H_{\text{TB}} + V \sum_{\langle i,j \rangle} \left(n_i-\frac{1}{2}\right) \left(n_j-\frac{1}{2}\right), \label{eq:tV_model}
\end{align}
where $H_{\mathrm{TB}}$ is the tight-binding Hamiltonian defined in Eq.~\eqref{eq:TB_model}, $n_{i} = c_{i}^{\dagger} c_{i}$ is the fermion number operator, and $V$ denotes the strength of the nearest-neighbor density-density interaction.
We set $V/t = 2$.

\begin{figure}
    \centering
    \includegraphics[width=\linewidth]{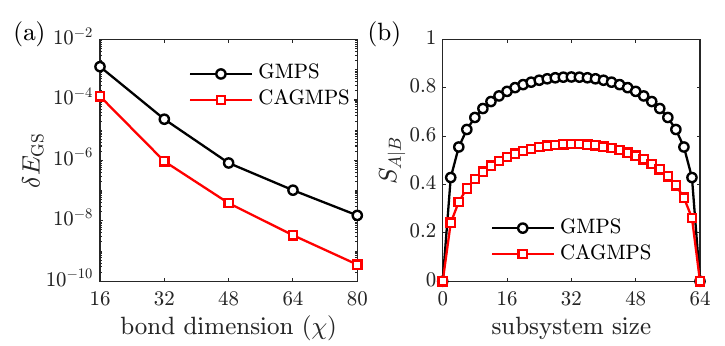}
    \caption{Benchmarks for the interacting spinless fermion $t$--$V$ model.
    (a) Ground-state energy errors of GMPS and CAGMPS for the $t$--$V$ model in Eq.~\eqref{eq:tV_model} versus bond dimension $\chi$ for $V/t = 2$ and $L = 64$.
    The reference energy is obtained from a high-bond-dimension DMRG calculation with $\chi = 1000$.
    (b) Bipartite entanglement entropy across different cuts for the same system at $\chi = 80$.}
    \label{fig:tV}
\end{figure}

Figure~\ref{fig:tV}(a) shows the ground-state energy error for $L = 64$, using a high-bond-dimension DMRG calculation with $\chi = 1000$ as the reference.
CAGMPS again gives lower errors than GMPS at fixed bond dimension, demonstrating that the improvement is not restricted to the free-fermion case.
The entanglement profiles in Fig.~\ref{fig:tV}(b) show the same tendency as in Fig.~\ref{fig:TB}(b).
For $L = 64$ and $\chi = 80$, CAGMPS has lower bipartite entanglement entropy than GMPS across the chain.
This indicates that the Clifford disentangling step remains effective in the presence of nearest-neighbor density-density interactions.

Finally, we study the spinless fermion $t$--$V$--$V'$ model,
\begin{align}
    H_{t\text{--}V\text{--}V'} &= H_{t\text{--}V} + V' \sum_{\langle \hspace{-0.2em} \langle i,j \rangle \hspace{-0.2em} \rangle} \left(n_i-\frac{1}{2}\right) \left(n_j-\frac{1}{2}\right), \label{eq:tVVp_model}
\end{align}
where $H_{t\text{--}V}$ is the $t$--$V$ model Hamiltonian defined in Eq.~\eqref{eq:tV_model}, $\langle \hspace{-0.2em} \langle i,j \rangle \hspace{-0.2em} \rangle$ denotes next-nearest-neighbor pairs, and $V'$ is the next-nearest-neighbor density-density interaction strength.
We choose $V/t = 4$, and $V'/t = 3.5$.

\begin{figure}
    \centering
    \includegraphics[width=\linewidth]{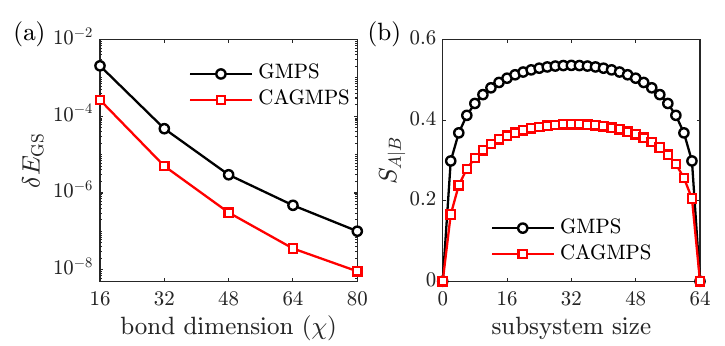}
    \caption{Benchmarks for the spinless fermion $t$--$V$--$V'$ model with nearest- and next-nearest-neighbor density-density interactions.
    (a) Ground-state energy errors of GMPS and CAGMPS for the $t$--$V$--$V'$ model in Eq.~\eqref{eq:tVVp_model} as a function of bond dimension $\chi$, at $V/t = 4$, $V'/t = 3.5$, and $L = 64$.
    The reference energy is obtained from a DMRG calculation with $\chi = 1000$.
    (b) Bipartite entanglement entropy of the same model across the chain, evaluated for $L = 64$ and $\chi = 80$.}
    \label{fig:tVVp}
\end{figure}

As shown in Fig.~\ref{fig:tVVp}(a), CAGMPS produces lower ground-state energy errors than GMPS when compared with the $\chi = 1000$ DMRG reference.
The corresponding entanglement entropy in Fig.~\ref{fig:tVVp}(b) is also reduced throughout the chain.
These results show that the advantage of CAGMPS persists even when longer-range density-density interactions are included.

These benchmark results highlight the general applicability and efficiency of Clifford augmentation in the GMPS framework.
Improvements over GMPS are observed not only for the free-fermion chain, but also for interacting models and a 2D lattice benchmark, indicating that the local Clifford transformations provide an effective compression mechanism for fermionic TN states. 
The observed improvements in accuracy and entanglement reduction, together with the preservation of fermionic locality and the compact parity-constrained Clifford gate set, make CAGMPS a promising variational ansatz for strongly correlated fermionic systems.

\section{Discussion}\label{sec:Discussion}
In this work, we have proposed a DMRG framework based on the CAGMPS ansatz, in which local Clifford circuits are embedded directly within the fermionic Grassmann formalism. 
This framework preserves fermionic locality without resorting to JW mappings, guarantees disentangling operations consistent with fermionic statistics, and substantially improves approximation accuracy. 
Moreover, owing to fermion-parity conservation and equivalence under entangling action, the relevant two-site Clifford gates are restricted to 12 inequivalent representatives, enabling a more efficient search than in the qubit setting.
As a result, the proposed approach provides a transparent, scalable, and effective variational approach for simulating strongly correlated fermionic systems.

The present framework also opens several promising directions for further research.
A natural extension is to higher-dimensional settings, where GTNs can be generalized to fermionic PEPS and related architectures~\cite{Corboz2009fermionic,Corboz2010simulation}. 
Since the Clifford disentanglers are formulated directly in the fermionic Grassmann language, embedding them into two-dimensional fermionic TNs may provide both conceptual clarity and computational advantages.
This offers a promising route toward simulating strongly correlated fermionic systems beyond one dimension.

Beyond ground-state simulations, the Clifford-augmented MPS has also been applied to the compression of quantum circuits in qubit and qutrit systems~\cite{fux2025disentangling,harper2025gcamps}. 
In particular, deep quantum circuits consisting mostly of Clifford gates with a limited number of non-Clifford gates can often be compressed efficiently when the non-Clifford content remains sufficiently sparse.
This can be understood by absorbing the non-Clifford resources into a modified input state, for example, by replacing selected qubits with magic states, so that the remaining evolution is described by a Clifford circuit.
From a broader perspective, it would be interesting to explore whether analogous circuit-compression schemes can be realized in fermion-based quantum simulations using CAGMPS.

Finally, it would be interesting to clarify the physical meaning of the entanglement reduction induced by fermionic Clifford circuits.
Such reductions may be related to changes in boundary conditions and duality transformations~\cite{Fan2025disentangling}, providing a useful direction for further investigation.
In critical systems, these effects may be analytically characterized using tools from boundary conformal field theory~\cite{hoshino2025stabilizer}.
Moreover, this perspective highlights the need to understand fermionic Clifford circuits not only as algorithmic building blocks, but also as resources whose power should be characterized relative to fermionic magic~\cite{Collura2024quantum,Sierant2025fermionic}. 
Clarifying this connection would be important for assessing classical simulability and for placing the efficiency of the method within a broader resource-theoretic framework.

\bigskip
\section*{acknowledgments}
We thank Chia-Min Chung, Marcello Dalmonte, Yoshiki Fukusumi, Masahiro Hoshino, Yi-Ping Huang, Hosho Katsura, Minsoo L. Kim, Tomotaka Kuwahara, Seung-Sup B. Lee, Mingpu Qin, Luca Tagliacozzo, Hung-Hsuan Teh, and Xhek Turkeshi for their insightful discussions and helpful comments.
W.-L. T. and T. O were supported by the Center of Innovation for Sustainable Quantum AI (JST Grant Number JPMJPF2221).
W.-L. T. was supported by JSPS KAKENHI Grant Numbers JP25H01545 and JP26K17054.
A. Y. was also supported by JST-CREST JPMJCR24I3.
K. O. was supported by KAKENHI Grants, Nos. JP21H05182 and JP21H05191, as well as JST-CREST No. JPMJCR24I1.
T. O was supported by KAKENHI Grant, Nos. 23H03818 and 22K18682.
D. K. was supported by the RIKEN Special Postdoctoral Researcher Program, the RIKEN Hakubi Project, and JSPS KAKENHI Grant Number JP26K17060.
The authors thank the Sustainable Quantum AI in Japan and the Physics Division of the National Center for Theoretical Sciences in Taiwan, where part of this work was carried out during the ``SQAI-NCTS Workshop on Quantum Technologies and Machine Learning." The authors also thank the Yukawa Institute for Theoretical Physics at Kyoto University, where part of this work was conducted during the program YITP-I-25-02, ``Recent Developments and Challenges in Tensor Networks: Algorithms, Applications to Science, and Rigorous Theories."

\section*{data availability}
We plan to publish a Grassmann tensor network toolbox \texttt{GrassmannTN2.jl} publicly at a later date. The scripts for performing the CAGMPS will not be publicly distributed due to a lack of general distribution formatting, but can be shared privately upon reasonable request.

\bibliography{draft}

\appendix
\section{Twelve Entangling Clifford Circuits}\label{App:List}

The 12 independent Clifford gates are listed as follows:
$$
\begin{array}{lll}
\text{\textbullet}\;\mathcal{I}
& & \text{\textbullet}\;\CNOT{01}\Sgate{1}\CNOT{01}
\\
\text{\textbullet}\;\CNOT{10}\CNOT{01}\CNOT{10}
& & \text{\textbullet}\;\Sgate{0}\CNOT{01}\Hgate{0}\CNOT{01}
\\
\text{\textbullet}\;\Hgate{0}\CNOT{01}\CNOT{10}\Hgate{1}
& & \text{\textbullet}\;\Sgate{0}\CNOT{10}\Hgate{1}\CNOT{10}
\\
\text{\textbullet}\;\CNOT{10}\Sgate{1}\CNOT{01}\Hgate{1}\CNOT{10}
& & \text{\textbullet}\;\CNOT{01}\Sgate{0}\CNOT{10}\Hgate{0}\CNOT{01}
\\
\text{\textbullet}\;\CNOT{01}\Sgate{0}\Hgate{0}\Sgate{0}\CNOT{01}
& & \text{\textbullet}\;\Sgate{0}\CNOT{10}\Hgate{1}\CNOT{10}\Sgate{0}
\\
\text{\textbullet}\;\Sgate{1}\CNOT{01}\Hgate{0}\CNOT{10}\Sgate{0}\CNOT{01}
& & \text{\textbullet}\;\CNOT{01}\CNOT{10}\Sgate{0}\Hgate{0}\Sgate{0}\CNOT{01}
\end{array}
$$
Here, $\mathcal{I}$ is the 2-site identity gate, $\Hgate{0/1}$ is the Grassmann Hadamard gate on the 1$^\text{st}$/2$^\text{nd}$ site, $\Sgate{0/1}$ is the Grassmann $S$-gate on the 1$^\text{st}$/2$^\text{nd}$ site, and $\CNOT{ij}$ is the Grassamnn CNOT-gate with controlled bit $i$ on the target bit $j$. To construct these generator gates, we first write the non-Grassmann gates as $4\times4$ matrices:
\begin{align}
    \text{I}&=\mat{1&&&\\&1&&\\&&1&\\&&&1},\\
    \text{S}^0&=\mat{1&&&\\&1&&\\&&i&\\&&&i},\;\text{S}^1=\mat{1&&&\\&i&&\\&&1&\\&&&i},\\
    \text{H}^0=\tfrac{1}{\sqrt{2}}&\mat{
    1 & 0 & 1 & 0 \\
    0 & 1 & 0 & 1 \\
    1 & 0 & \!\!\!-1 & 0 \\
    0 & 1 & 0 & \!\!\!-1},\;
    \text{H}^1=\tfrac{1}{\sqrt{2}}\mat{
    1 & 1 & 0 & 0 \\
    1 & \!\!\!-1 & 0 & 0 \\
    0 & 0 & 1 & 1 \\
    0 & 0 & 1 & \!\!\!-1},\\
    \text{CNOT}^{01}=&\mat{
    1 & 0 & 0 & 0 \\
    0 & 1 & 0 & 0 \\
    0 & 0 & 0 & 1 \\
    0 & 0 & 1 & 0},\;
    \text{CNOT}^{10}=\mat{
    1 & 0 & 0 & 0 \\
    0 & 0 & 0 & 1 \\
    0 & 0 & 1 & 0 \\
    0 & 1 & 0 & 0}.
\end{align}
The corresponding Grassmann matrices are then constructed via
\begin{equation}
    \mathcal{M}_{\psi^\dagger\phi}=\sum_{IJ} M_{IJ}\psi^{\dagger I}\phi^J
\end{equation}
where $I,J$ run from 0 to 3 and $M$ is one of the $4\times 4$ matrices given above. We then split the fermions $\psi^\dagger$ and $\phi$ into two components via the algebra splitting procedure described in Appendix A.3 of Ref \cite{Yosprakob2023grassmanntn}, thus obtaining a $2\times2\times2\times2$ tensor. Note that because of the sign factors introduced by the algebra splitting, the 4-legged Grassmann tensors cannot be imported directly with the 4-legged non-Grassmann tensors as the coefficients.

\end{document}